\documentclass {aa}
\usepackage{natbib,graphicx,hyperref} 
\usepackage[varg]{txfonts}

\begin{document}

\title{No evidence for synchronization of the solar cycle by a ``clock''} 

\author{ E. Weisshaar\inst{1} \and R.~H. Cameron\inst{2}
  \and M. Sch{\"u}ssler\inst{2} } 

\institute {Brunnenstr. 58, 61231 Bad Nauheim, Germany
  \and
  Max-Planck-Institut f{\"u}r Sonnensystemforschung,
  Justus-von-Liebig-Weg 3, 37077 G{\"o}ttingen,
  Germany\\ 
  \email{cameron@mps.mpg.de} }
 
\date{Received ; accepted}

\abstract{The length of the solar activity cycle fluctuates
  considerably. The temporal evolution of the corresponding cycle
  phase, that is, the deviation of the epochs of activity minima or
  maxima from strict periodicity, provides relevant information
  concerning the physical mechanism underlying the cyclic magnetic
  activity. An underlying strictly periodic process (akin to a perfect
  ``clock''), with the observer seeing a superposition of the perfect
  clock and a small random phase perturbation, leads to long-term
  phase stability in the observations.  Such behavior would be
  expected if cycles were synchronized by tides caused by orbiting
  planets or by a hypothetical torsional oscillation in the solar
  radiative interior. Alternatively, in the absence of such
  synchronization, phase fluctuations accumulate and a random walk of
  the phase ensues, which is a typical property of randomly perturbed
  dynamo models. Based on the sunspot record and the reconstruction of
  solar cycles from cosmogenic $^{14}$C, we carried out rigorous
  statistical tests in order to decipher whether there exists phase
  synchronization or random walk. Synchronization is rejected at
  significance levels of between 95\% (28 cycles from sunspot data)
  and beyond 99\% (84 cycles reconstructed from $^{14}$C), while the
  existence of random walk in the phases is consistent with all data
  sets. This result strongly supports randomly perturbed dynamo models
  with little inter-cycle memory.}


\authorrunning{Weisshaar et al.}

\titlerunning{No evidence for synchronization of the solar cycle} 

\maketitle

\section{Introduction}
\label{sec_intro}
The historical sunspot record reveals significant variability in the
amplitude and length of the individual activity cycles.  In order to
understand the physical mechanism underlying the variable cyclic
activity, it is important to answer the question of whether the phase
of the activity cycle shows long-term stability. In this case,
although the length of the individual cycles fluctuates, the cycle
appears to remain synchronized to some strictly periodic process (akin
to a ``clock''). This means that the fluctuations of the cycle length
do not correspond to real phase changes but reflect, for instance,
observational errors or varying transit times of magnetic flux rising
through the convection zone. Such a situation would occur if the
activity cycles were triggered, for example, by a hypothetical high-Q
torsional oscillation in the solar interior \citep{Dicke:1970} or by
minuscule tidal forces exerted by the planets
\citep[e.g.,][]{Stefani:etal:2021}. Alternatively, in the absence of
such synchronization in the system, phase fluctuations can lead to a
random walk of the phase.  This would happen in the case of a cyclic
dynamo with a period that varies because of random fluctuations of
dynamo excitation or meridional flow. We note that the clock case
could also correspond to a dynamo, albeit one with an externally
enforced fixed period.
 
\citep{Yule:1927} took an ideal pendulum as a simple mechanical example
to illustrate both cases: if the motion of the pendulum is
undisturbed, but the observation of its position is subject to
measurement error, we have a case of phase stability and
synchronization. Alternatively, if the pendulum were to be randomly
bombarded with peas, the resulting phase fluctuations would accumulate
and the phase would show a random walk.

There have been previous attempts in the literature to explore the phase
stability of the solar cycle on the basis of empirical data. The
sunspot record since 1700 provides a relatively consistent dataset,
but has generally been considered to be too short for a reliable
statistical phase analysis \citep[e.g.,][]{Hoyng:1996}. Further
extension of the cycle into the past has been attempted on the basis
of reports of pre-telescopic sightings of sunspots and aurorae.  The
problem there is that the sporadic and often ambiguous reports are
much too sparse to faithfully determine the times of maxima and minima
of individual cycles with yearly precision
\citep[e.g.,][]{Stephenson:1988, Stephenson:1990}. Nevertheless,
\citet{Schove:1955,Schove:1983} gave years of sunspot minima and
maxima back to the year 649 B.C., mainly relying on aurora
reports. However, as he imposed a fixed number of nine cycles per
century, he assumed phase stability beforehand \citep{Nataf:2022}, meaning
that these data are useless for our purposes.  Catalogs of naked-eye
observations of sunspots in the pre-telescopic era
\citep{Wittmann:1978,Wittmann:Xu:1987} are even more sparse and show
no clustering around the maxima of a hypothetical phase-stable cycle
\citep{Stix:1983, Stix:1984}.  Furthermore, \citet{Carrasco:etal:2020}
compared naked-eye observations with telescopic records in the 19th
century and found that the naked-eye observations do not necessarily
correspond to high solar activity or large sunspot groups, meaning that such
observations cannot be used offhandedly to infer the timing of
individual cycle maxima \citep[see also][]{Usoskin:etal:2015}. Similar
arguments apply to historical reports considered to represent aurorae,
although long-term activity variations and grand minima may
tentatively be identified in these data. A comprehensive and critical
summary of the use of historical documents to infer solar phenomena in
the past was given by \citet{Vaquero:Vazquez:2009}.

\begin{figure}
\centering
\includegraphics[width=\hsize]{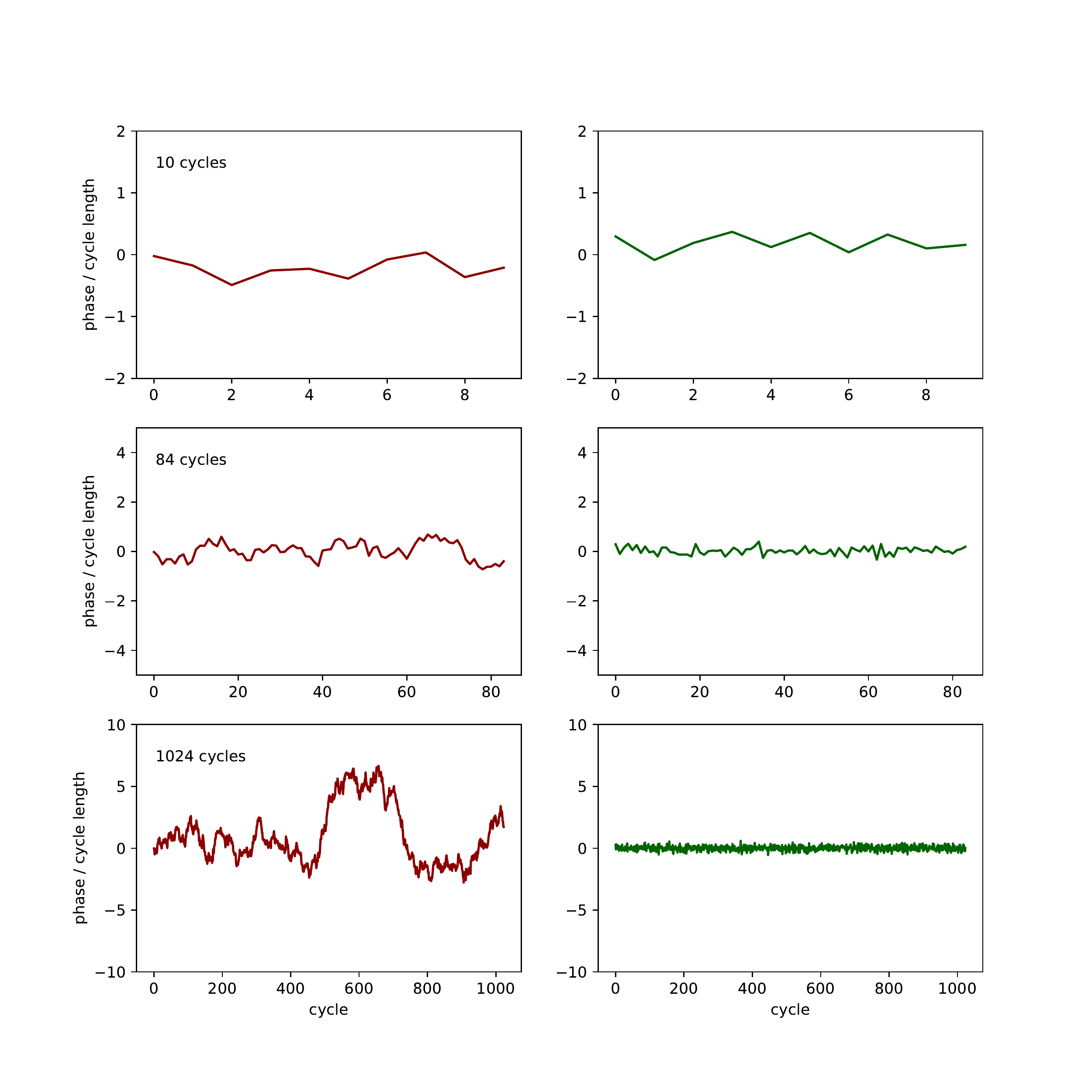}
\caption{Examples of Monte Carlo simulations of the phase evolution
  for case R (random walk, left panels) and for case C
  (synchronization, right panels) and different numbers of cycles.}
\label{fig:phases}
\end{figure}

Although claims of phase stability of the solar cycle can be found in
various papers \citep[e.g.,][]{Lomb:2013, Russell:etal:2019,
  Stefani:etal:2020}, a conclusive answer to the question of phase
stability or otherwise requires both a clear definition of phase
stability and proper statistical analysis. For instance, nonmonotonic
evolution of phase in the sense that sequences of positive excursions
of phase over a number of cycles are seemingly ``compensated for'' by
later cycles \citep[e.g.,][]{Charbonneau:Dikpati:2000} does not
necessarily entail phase stability, synchronization, or memory (see
examples in Fig.~\ref{fig:phases}). This refers also, for instance, to
the well-known weak anti-correlation between cycle length and amplitude
\citep{Hoyng:1996,Lomb:2013}.

The first systematic statistical studies of the phase stability of the
solar cycle were presented by \citet{Dicke:1978} and
\citet{Gough:1978}, who carried out conceptually similar analyses
based on the telescopic sunspot record. Both authors concluded that
the available data were not sufficient to decisively discriminate
between random walk and phase synchronization.  Subsequently,
\citet{Gough:1981, Gough:1983, Gough:1988} corrected and modified his
earlier analysis, still concluding that the results do not permit
the rejection of one of the alternatives.

In this paper, we revisit the problem using the additional cycle
maximum and minimum epochs until 2019 (28 cycles compared to 24 cycles
available in 1981) as well as the reconstruction of 84 solar cycles
between the years 976 and 1888 based on a new analysis of $^{14}$C
in tree rings \citep{Brehm:etal:2021, Usoskin:etal:2021}. Extending
the analyses of \citet{Gough:1981, Gough:1983}, we find that these data
are statistically consistent with random walk of the cycle phase and
exclude clock synchronization with at high levels of significance.

The paper is organized as follows. In Sect.~\ref{sec_Gough} we consider
the disparate phase evolution of a randomly perturbed, memory-less
oscillator (random walk of phase) and of a perturbed
clock-synchronized system and reconsider the statistical method of
\citet{Gough:1981, Gough:1983}, thereby correcting an error in his
original analysis. We apply the method to the up-to-date sunspot
record and to the $^{14}$C-based reconstruction in
Sect.~\ref{sec_results}. Further tests concerning the significance of
the results are presented in Sect.~\ref{sec_tests}. 
Section~\ref{sec_concl} contains our conclusions.

\section{Random walk of phase versus synchronization: statistical analysis}
\label{sec_Gough}
\subsection{The method of Gough(1981)}
\label{subsec_Gough_1981}
Following \citet{Gough:1981}, we consider two extreme cases as the
basis of our statistical analysis. The first case corresponds to the
absence of inter-cycle memory: stochastic fluctuations of cycle length
lead to random walk of phase. The alternative case is a perturbed
cyclic system whose period is fixed through synchronization by a
perfect clock: any fluctuation in the timing of cycle number $n$ does
not affect the timing of cycle $n+1$. For convenience, we call the
first model {\it random walk\/} (in short: case R) and the second model
{\it clock\/} (in short: case C).

We consider epochs of the cycle minima (or maxima), $t_n$ ($n=0,\dots
N$).  In both cases (R and C), we have random fluctuations of the
cycle length, for example due to its dependence on fluctuating parameters
(case R) or to time delays of the transfer of magnetic flux
through the turbulent convection zone (case C).  In case C, the
resulting phase fluctuations do not accumulate, because the governing
periodic process runs with a fixed period and phase, meaning that previous
fluctuations appear to be ``corrected'' by synchronization, and we have
\begin{equation}
  t_{n,\mathrm C} = n\cdot P_{\mathrm C} + \tau_n\, , 
\label{eq:epochs_clock}   
\end{equation}
where $P_{\mathrm C}$ is the clock period and $\tau_n$ is the (random) phase
fluctuation in cycle $n$. In contrast, for case R, the phase
fluctuations, $\psi_i$, accumulate, so that
\begin{equation}
  t_{n,\mathrm R} = n\cdot P_{\mathrm R} + \sum_{i=0}^n \psi_i\, ,
\label{eq:epochs_dynamo}       
\end{equation}
where $P_{\mathrm R}$ is the period at which the undisturbed system
would operate in the absence of fluctations. In practice, both
  $P_{\mathrm C}$ and $P_{\mathrm R}$ correspond to the average of the
  cycle length over an infinite set of cycles, so that, in principle,
  they could have been denoted by the same symbol. For clarity, we
  continue to use separate symbols in what follows. These two simple models
confer an advantage  in that they are quite general and independent of the
detailed nature of the phase changes (e.g., discrete random jumps or
more gradual variations during the cycle). Of course, more explicit
models could be analyzed more deeply and therefore possibly yield more
detailed results.

We illustrate
the disparate phase evolution for both cases using a simple example.  We
assume $P_{\mathrm R} = P_{\mathrm C} = 11\,$yr and normally
distributed phase fluctuations with zero mean and a standard deviation
of $2\,$yr. Figure~\ref{fig:phases} shows the evolution of the phases
for both cases ($\tau_n$ and $\sum\psi_i$, respectively) and for
different total numbers of cycles (note the different scaling of the
ordinate axes). The two cases are virtually indistinguishable if only
ten periods are considered.  With 84 cycles, which is the length of the
reconstructed sunspot record by \citet{Usoskin:etal:2021}, differences
are visible and finally become obvious for 1024 cycles. We note that
random walk of phase does not imply that the phase evolves
monotonically; indeed in our case of a probability distribution for
phase fluctuations that is symmetric with respect to zero, a random
walk corresponds to a recurrent Markov chain that returns to zero
infinitely often. Such returns have therefore nothing to do with any
kind of ``clock correction of phase deviations''. The random walk with
1024 cycles clearly shows such returns as well as intervals with
approximately linear phase drift, which, if viewed in isolation, could
easily be mistaken as evidence for phase stability with respect to a
matching period.

In the case of the solar cycle, available datasets are not sufficiently
long to decide the case simply by visual inspection of the phase evolution:
a proper statistical analysis needs to be carried out. Here, we follow the
approach of \citet{Gough:1981, Gough:1983}, which is sketched
below (an example of the detailed calculations is given in the Appendix).

We have a series of epochs of cycle minima (or maxima), $t_n$ $(n=0,
\dots N)$, covering $N$ cycles. The aim of the analysis is to define
statistical criteria that allow us to decide whether these data are
consistent with (or exclude) the clock case (C, synchronization)
or random phase walk (case R) as given in Eqs.~(\ref{eq:epochs_clock})
and (\ref{eq:epochs_dynamo}). The observed period of cycle $n$, $P_n =
t_n - t_{n-1}$, is given by
\begin{equation}
  P_{n,{\mathrm C}} = P_{\mathrm C} + \tau_n - \tau_{n-1}  
\label{eq:period_clock}  
\end{equation}
for case C and by
\begin{equation}
  P_{n,{\mathrm R}} = P_{\mathrm R} + \psi_n
\label{eq:period_dynamo}
\end{equation}
for case R. As the ``true'' periods, $P_{\mathrm C}$ and $P_{\mathrm R}$, are
not known from the data, they are estimated by the mean period over
the $N$ empirical cycles, $\langle P \rangle = (t_N - t_0)/N$, meaning that we have
\begin{equation}
  \langle P \rangle{_{\mathrm C}} = P_{\mathrm C} + {1\over N}(\tau_N-\tau_0)
\label{eq:mean_period_clock}
\end{equation}
for case C and
\begin{equation}
  \langle P \rangle{_{\mathrm R}} = P_{\mathrm R} + {1\over N} \sum_{i=1}^N \psi_i   
\label{eq:mean_period_dynamo}
\end{equation}
for case R.
The phase deviations with respect to the mean period are generally
defined as
\begin{equation}
  \phi_n = t_n - t_0 - n\langle P \rangle \,,
\label{eq:phase_deviations}
\end{equation}
meaning that we obtain 
\begin{equation}
  \phi_{n,{\mathrm C}} = t_{n,\mathrm C} - t_{0,\mathrm C} - n\cdot \langle P \rangle_{\mathrm C} = \tau_n - \tau_0 - {n\over N} (\tau_N - \tau_0) 
\label{eq:phases_clock}
\end{equation}
for case C, and
\begin{equation}
  \phi_{n,{\mathrm R}} = t_{n,\mathrm R} -  t_{0,\mathrm R} - n\cdot \langle P \rangle_{\mathrm R} =  \sum_{i=1}^n \psi_i -  {n\over N} \sum_{i=1}^N \psi_i 
\label{eq:phases_dynamo}
\end{equation}
for case R. In both cases, we therefore start with zero phase deviation for $n=0$.

The basis for the statistical analysis are the expectation values of
the variances of cycle period and phase deviation, namely
\begin{equation}
  \sigma{_{\mathrm P}}^2 = {1\over N}  \sum_{i=1}^N (P_i - \langle P \rangle)^2
\label{eq:variance_period}     
\end{equation}
for the period and
\begin{equation}
  \sigma{_\phi}^2 =  {1\over N+1}  \sum_{i=0}^N \left( \phi_i - \langle \phi \rangle \right)^2
   = {1\over N+1}  \sum_{i=0}^N\phi_i^2 - \left({1\over N+1}  \sum_{i=0}^N\phi_i \right)^2     
\label{eq:variance_phase}
\end{equation}
for the phase deviations, where the angular brackets indicate the
arithmetic mean. For both cases, we assume uncorrelated phase
fluctuations ($\tau_i$ and $\psi_i$, respectively) with zero mean.  A
somewhat lengthy, but straightforward calculation (see Appendix)
yields the expectation values of the variances as functions of the
number of cycles ($N$) and the amplitudes of the fluctuations
($\mathcal{T}^2$ and $\Psi^2$, respectively).  For the expectation value
of the variance of period, we obtain
\begin{equation}
  E(\sigma_{\mathrm P}^2)_{\mathrm C} = 2 \left( 1 - {1\over N^2}\right)\mathcal{T}^2
\label{eq:exp_period_clock}    
\end{equation}
for case C and
\begin{equation}
  E(\sigma_{\mathrm P}^2)_{\mathrm R}= {N-1\over N}\Psi^2
\label{eq:exp_period_dynamo}     
\end{equation}
for case R, both in agreement with \citet{Gough:1981}.
The expectation value of the variance of phase for case C is given by
\begin{eqnarray}
  E(\sigma_\phi^2)_{\mathrm C} &=& E\left({1\over N+1}\sum_{i=0}^N\phi_i^2\right) -  E\left(\left[{1\over N+1}\sum_{i=0}^N\phi_i \right]^2\right) \nonumber \\     
                      &=& {5N^2-6N+1\over 3N(N+1)}\mathcal{T}^2 - {(N-1)\over 2(N+1)}\mathcal{T}^2 \nonumber \\ 
                      &=& {(7N-2)(N-1)\over 6N(N+1)}\mathcal{T}^2 \,. 
\label{eq:exp_phase_clock}
\end{eqnarray}
Here, \citet{Gough:1981} apparently omitted the second term on the r.h.s.
of Eq.~(\ref{eq:variance_phase}), meaning that our result differs from his.
Likewise,  for case R, we obtain 
\begin{equation}
E(\sigma_\phi^2)_{\mathrm R} = {N-1\over 6}\Psi^2 - {N(N-1)\over 12(N+1)}\Psi^2 = {(N+2)(N-1)\over 12(N+1)}\Psi^2 \,. 
\label{eq:exp_phase_dynamo}
\end{equation}
The final statistic is defined as the ratio of the variances, meaning that
the unknown fluctuation amplitudes cancel out. The result is
\begin{equation}
S_{\mathrm C} = {E(\sigma_\phi^2)_{\mathrm C}\over E(\sigma_{\mathrm P}^2)_{\mathrm C}} = {N(7N-2)\over 12(N+1)^2}
\label{eq:exp_ratio_clock}
\end{equation}
for case C and
\begin{equation}
S_{\mathrm R} = {E(\sigma_\phi^2)_{\mathrm R}\over E(\sigma_{\mathrm P}^2)_{\mathrm R}} = {N(N+2)\over 12(N+1)}
\label{eq:exp_ratio_dynamo} 
\end{equation}
for case R. For large values of $N$, $S_{\mathrm C}$ approaches a constant value
of $7/12,$ while $S_{\mathrm R}$ approaches linear growth corresponding to
$N/12$. This is half of the slope obtained by \citet{Gough:1981}. 
We cross-checked the correctness of our analytical
calculations using Monte Carlo simulations.

\subsection{Modified method (Gough, 1983)}
\label{subsec_Gough_1983}
In follow-up papers, \citet{Gough:1983,Gough:1988} modified his
analysis, referring to \citet{Dicke:1978}. Instead of taking the
arithmetic mean of the cycle periods provided by the data as in
Eqs. \ref{eq:mean_period_clock} and \ref{eq:mean_period_dynamo}, an
estimate of the ``true'' periods, $P_{\mathrm C}$ or $P_{\mathrm R}$,
is determined by minimizing the variance of phase deviation,
$\sigma_\phi^2$. The phases are generally given by $\phi_n=t_n-t_0-nP$
and $P_{\rm min}$ is determined by the minimalization process, as
\begin{equation}
  P_{\rm min} = {6\over N(N+1)(N+2)}\sum_{i=0}^N(2i-N)t_i \,.
\label{eq:P_minimized}   
\end{equation}
The resulting periods for the two cases are then given by
\begin{equation}
  P_{\rm min}^{\mathrm C} = P_{\mathrm C} + {6\over N(N+1)(N+2)}\sum_{i=0}^N(2i-N)\tau_i 
\label{eq:P_minimized_C}   
\end{equation}
for case C, and
\begin{equation}
  P_{\rm min}^{\mathrm R} = P_{\mathrm R} +{6\over N(N+1)(N+2)}\sum_{i=0}^N(2i-N)\sum_{j=1}^N\psi_j 
\label{eq:P_minimized_D}   
\end{equation} 
for case R. Clearly, for growing values of $N$, these expressions
converge more rapidly to the ``true'' periods than the estimates
derived from the average period given by Eqs.~(\ref{eq:mean_period_clock})
and (\ref{eq:mean_period_dynamo}).

Based on the these period estimates, the corresponding ratios of expectation values
for the variances of phase deviation and period are given by
\begin{equation}
S_{\mathrm C} = {E(\sigma{_\phi}^2)_{\mathrm C}\over E(\sigma{_{\mathrm P}}^2)_{\mathrm C}} = {N^2\over 2(N+1)^2}
\label{eq:mod_exp_ratio_clock}
,\end{equation}
for case C and
\begin{equation}
S_{\mathrm R} = {E(\sigma{_\phi}^2)_{\mathrm R}\over E(\sigma{_{\mathrm P}}^2)_{\mathrm R}} = {N(N+3)\over 15(N+1)}
\label{eq:mod_exp_ratio_dynamo} 
\end{equation}
for case R, both in agreement with the expressions given by \citet{Gough:1983}.

\begin{figure}
\centering
\includegraphics[width=\hsize]{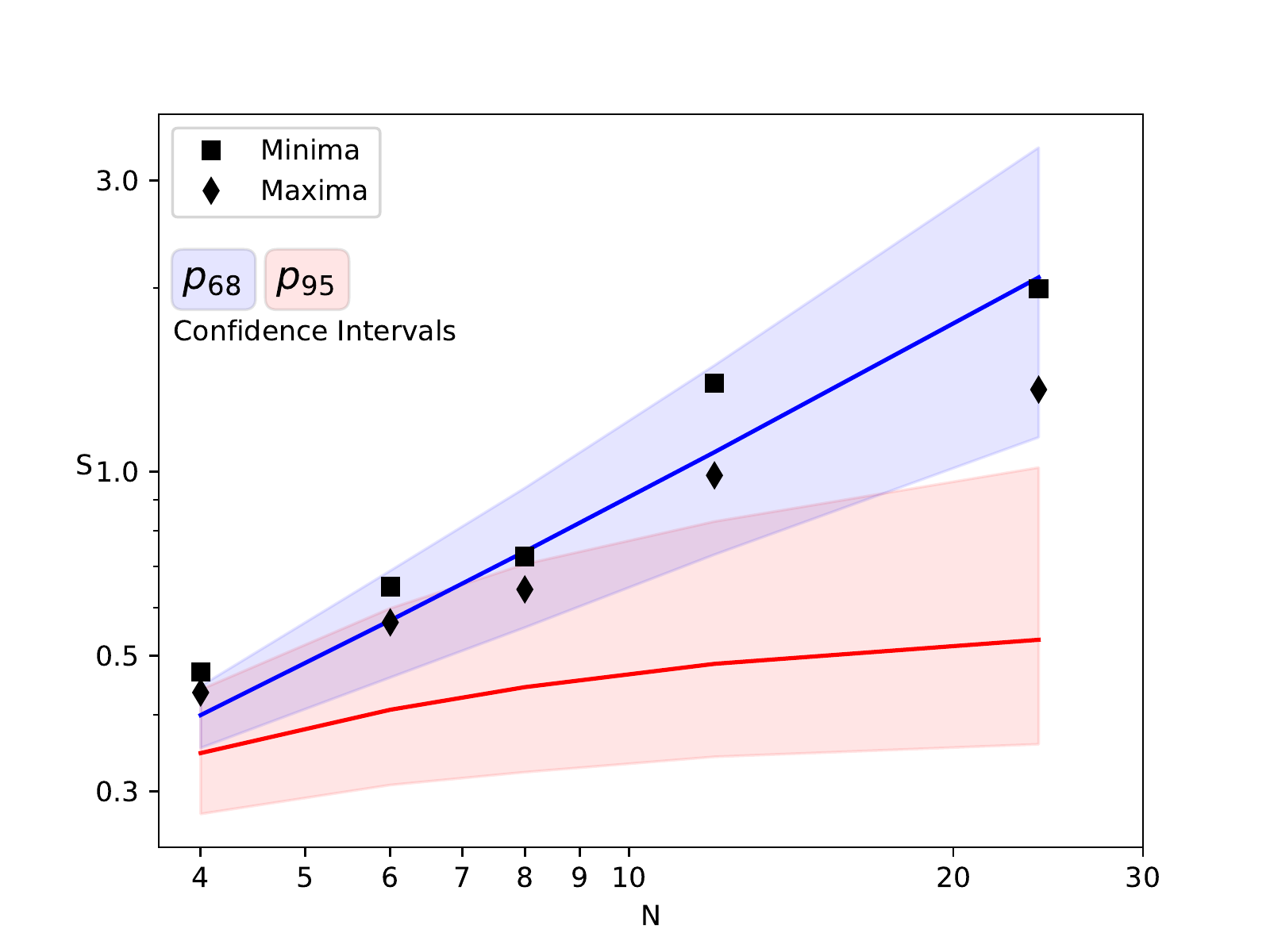}
\includegraphics[width=\hsize]{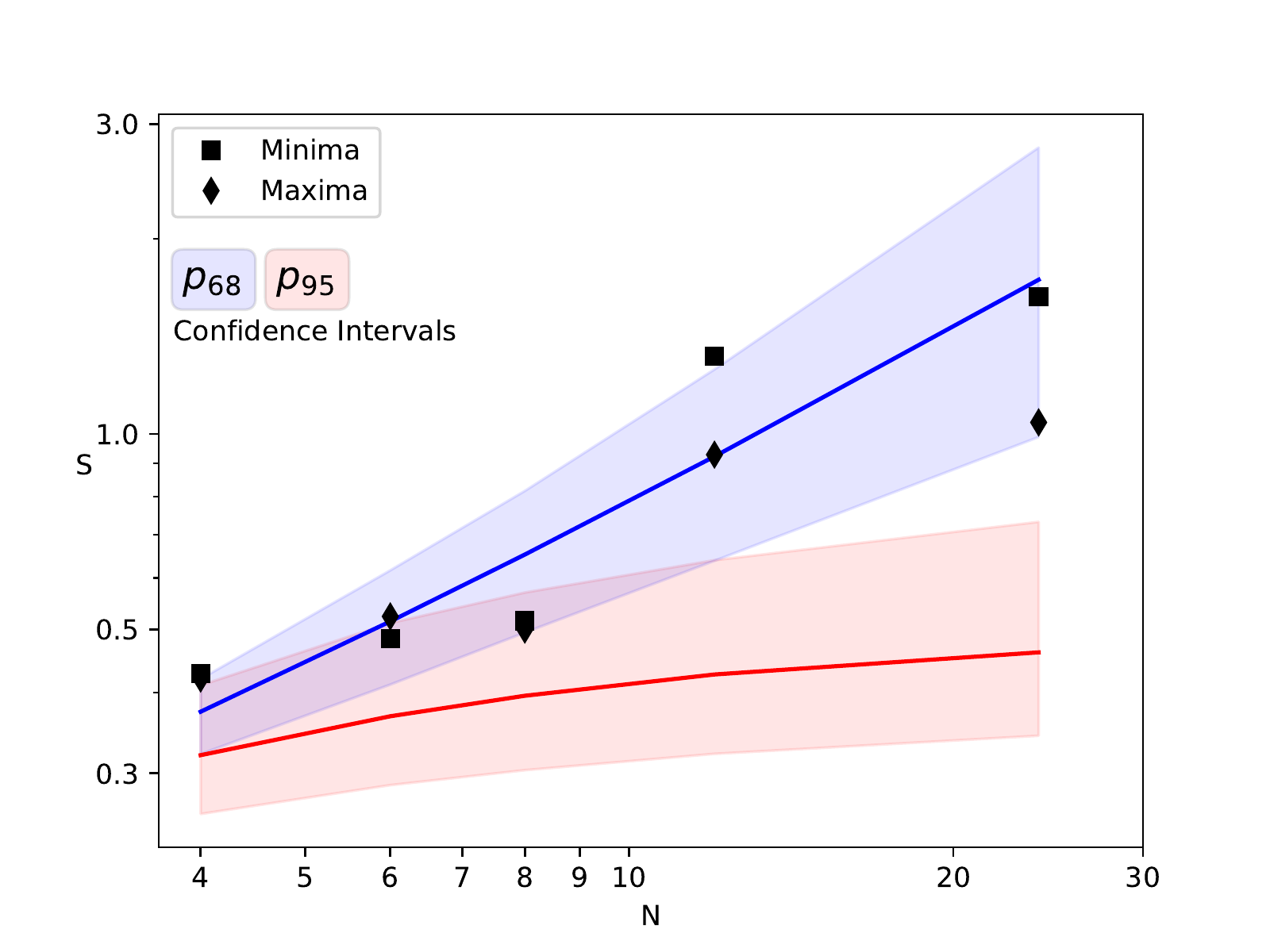}
\caption{Application of the statistical methods of \citet{Gough:1981,
    Gough:1983} to the epochs of sunspot minima and maxima of 24
  activity cycles between 1712 and 1976.  {\it Upper panel:}
  (Corrected) Original method of \citet{Gough:1981}. {\it Lower
    panel:} Modified method \citep{Gough:1983}.  In both panels, the
  symbols show the ratio
  $S=\langle\sigma_\phi^2\rangle/\langle\sigma_{\mathrm P}^2\rangle$
  as a function of $N$ (number of cycles in the data segments) for
  cycle minima (squares) and maxima (rhombs). The lines indicate the
  ratio of the expectation values, $E(\sigma_\phi^2)/E(\sigma_{\mathrm
    P}^2)$ for case R (random walk of phase; blue curve) and case C
  (clock synchronization; red curve), respectively. The shaded
  areas show one-sided 68\% $(p_{68})$ and 95\% $(p_{95})$ percentiles
  determined from 10 000 Monte-Carlo simulations for each case. In
  both panels, the data points align well with the curve corresponding
  to case R (random walk) while case C (synchronization) is rejected at a confidence level of 
  at least 95\%.}
\label{fig:sunspots_24}
\end{figure}

\begin{figure}
\begin{center}
\includegraphics[width=\hsize]{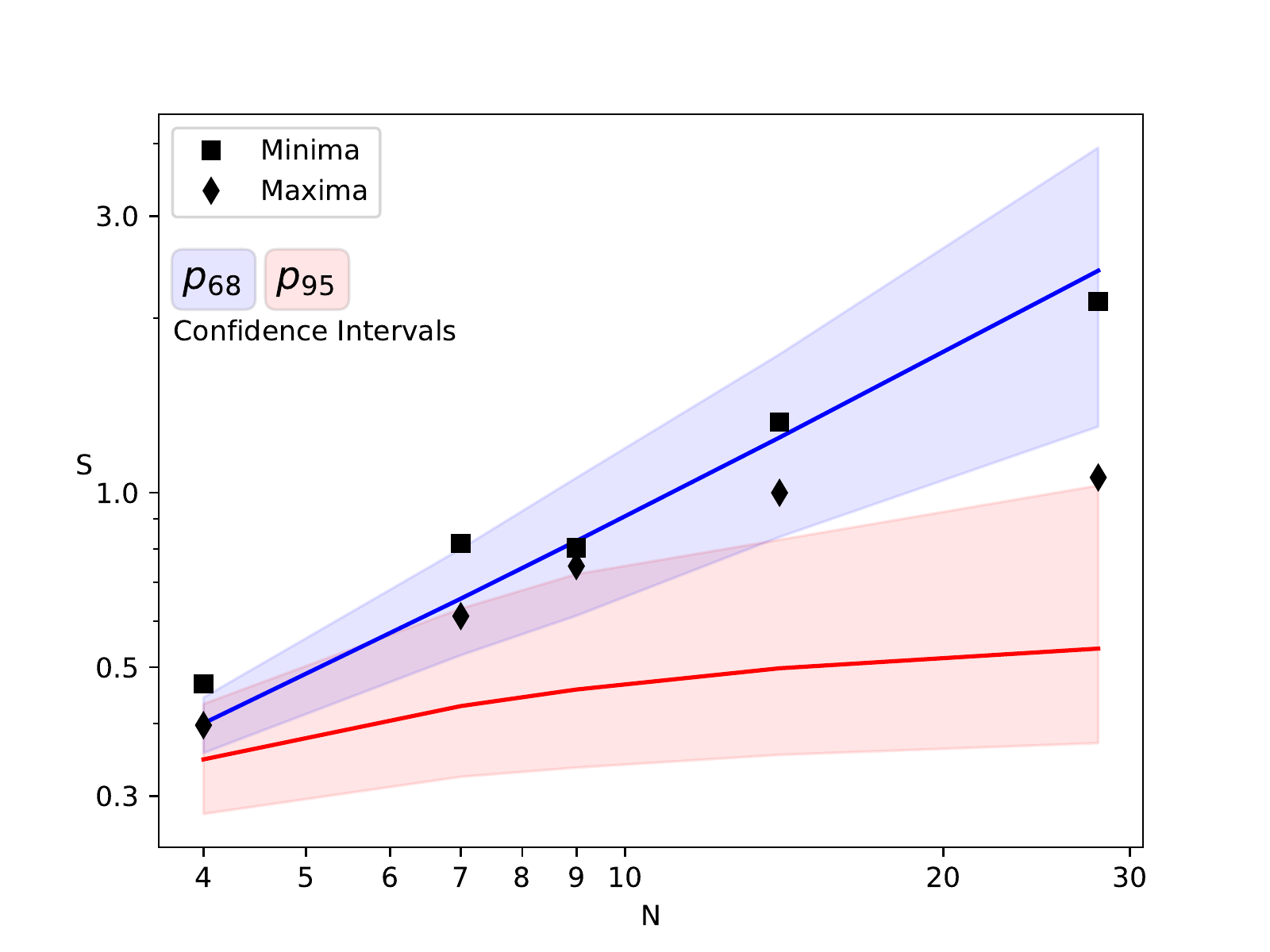}
\includegraphics[width=\hsize]{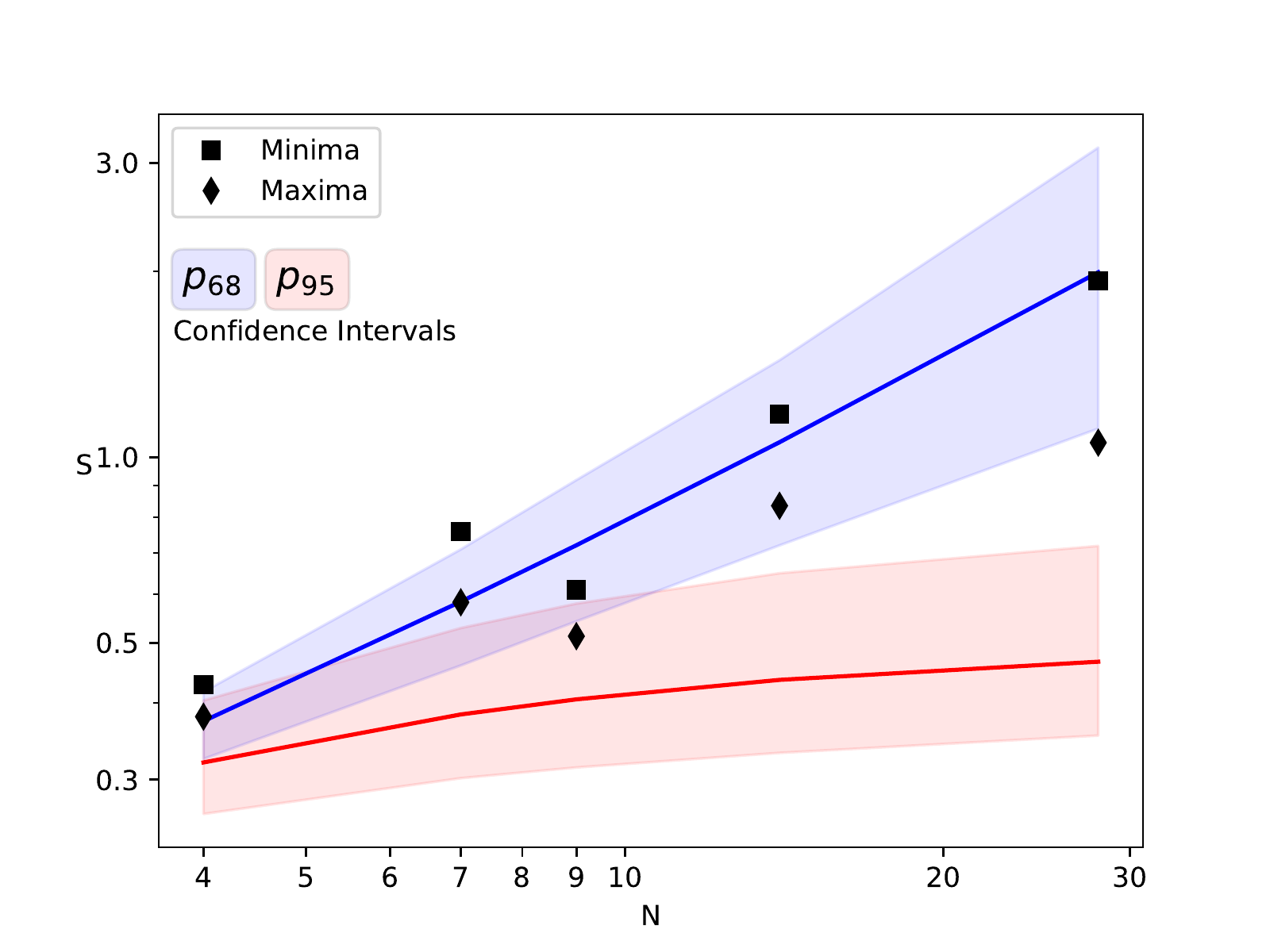}
\caption{Same as Fig.~\ref{fig:sunspots_24}, now on the basis of the
  sunspot data for 28 solar cycles between 1712 and 2019. Compared to
  Fig.~\ref{fig:sunspots_24}, the confidence intervals have become
  somewhat narrower, owing to the longer data set, thus strengthening the
  conclusion: clock synchronization is rejected at a confidence level
  of at least 95\%, while random walk of phase is consistent with the
  data.}
\label{fig:sunspots_28}
\end{center}
\end{figure}

\section{Application to observed and reconstructed sunspot cycles}
\label{sec_results}
We consider a set of empirical epochs of solar cycle minima (or
maxima), $t_n \, (n=0\dots M)$, corresponding to $M$ cycles. The
individual periods are then $P_i = t_i - t_{i-1}$ $(i=1\dots M)$ and
the phase deviations with respect to the estimate, $P_{\rm est}$, of
the ``true'' period are determined as $\phi_n = t_n - t_0 - n\cdot
P_{\rm est}\, (n=0\dots M)$.  $P_{\rm est}$ is calculated either as
the arithmetic mean of the individual periods (original Gough method)
or by minimalization of $\sigma_\phi^2$ (modified method).
 
The full data set of $M$ cycles is divided into contiguous segments of
$N(q)=M/q$ cycles each, where the values of $q$ are the divisors of
$M$. For each $q$, the mean variances, $\langle\sigma_{\mathrm
  P}^2\rangle$ and $\langle\sigma_\phi^2\rangle$, are determined
according to Eqs.~(\ref{eq:variance_period}) and
(\ref{eq:variance_phase}) as averages over the corresponding segments
of length $N(q)$. The ratio
$S=\langle\sigma_\phi^2\rangle/\langle\sigma_{\mathrm P}^2\rangle$ is
then plotted as a function of $N(q)$, to be compared with the
curves corresponding to the ratio of the analytically calculated
expectation values, $S_{\mathrm C}$ and $S_{\mathrm R}$.

Confidence intervals with regard to $S_{\mathrm C}$ and $S_{\mathrm R}$ are determined by Monte Carlo simulations. We assume normally
distributed phase fluctuations $\tau_n$ and $\psi_n$, respectively,
with zero mean and a standard deviation of 2~yr to compute
10 000~realizations of $M$ perturbed 11 yr cycles for each of the two
cases according to Eqs.~(\ref{eq:epochs_clock}) and
(\ref{eq:epochs_dynamo}).  The assumed standard deviation is
consistent with the value inferred for 84 solar cycles reconstructed
from $^{14}$C by \citet{Usoskin:etal:2021}, but the precise choice is
irrelevant, as the statistic $S$ neither depends on the
amplitude nor on the standard deviation of the distribution of the
phase fluctuations.  Following the procedures described in the
previous section, we then calculate the values of $S$ for each value
of $N(q)$ and for all realizations. From the resulting distributions
of $S_{\mathrm C}$ and $S_{\mathrm R}$, we determine the one-sided
68\%, 95\%, and 99\% percentiles: as the distributions
are generally not symmetric, we start from the average value and separately
count the number of realizations until we reach the given percentile
on each side. For a normal distribution, these percentiles would
(approximately) correspond to the $\pm1\sigma$, $\pm2\sigma$, and $\pm
3\sigma$ ranges, where $\sigma$ denotes the standard deviation.

As a first application, we consider the telescopic sunspot record.
\citet{Gough:1981, Gough:1983} considered the maximum and minimum
epochs of 24 cycles between the years 1712 and 1976. Using the epochs
provided by the SILSO database (www.sidc.be/silso), we redid
his analysis using both the (corrected) original method and the
modified method as presented in Sect.~\ref{sec_Gough}.  The results are
shown in Fig.~\ref{fig:sunspots_24}. Already with the 24 cycles
available to \citet{Gough:1981}, the data points for both variants of
the method align well with the blue line representing the expectation
values for case R.  Clock synchronization (case C) can be rejected at
the 95\% confidence level. Owing to its better convergence
property, the modified method provides somewhat narrower confidence
intervals than the original method.  The differences between the
diagram in the left panel of Fig.~\ref{fig:sunspots_24} and Fig.~7 of
\citet{Gough:1981} result from the correction of the calculation of
the variance of phase (see Sect.~\ref{sec_Gough}). This leads to a
downward shift of the theoretical curves for both cases. The result is
that even the limited data available in the early 1980s clearly favor
case R,that is, a random walk of the phase of the solar cycle. This is in
accordance with the results presented in \citet{Gough:1983} and
\citet{Gough:1988}.

In the meantime, data for 28 cycles have become available.
Figure~\ref{fig:sunspots_28} shows the results of the statistical
analysis. The longer data series confirms and strengthens the results
obtained on the basis of 24 cycles.  In particular, the confidence
intervals have become narrower compared to Fig.~\ref{fig:sunspots_24},
meaning that the rejection of the synchronization hypothesis becomes even
more significant.

\begin{figure}
\begin{center}
\includegraphics[width=\hsize]{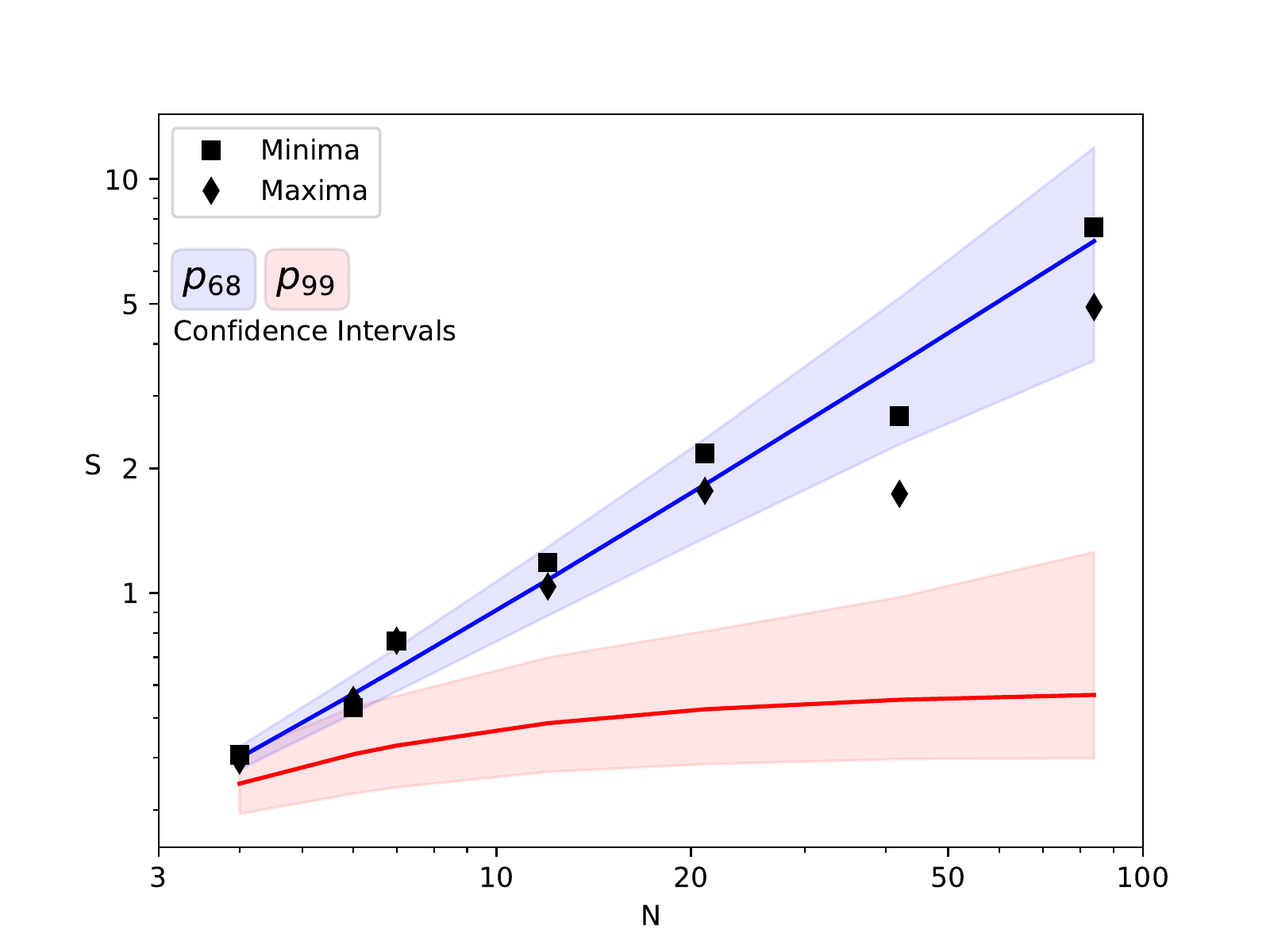}
\includegraphics[width=\hsize]{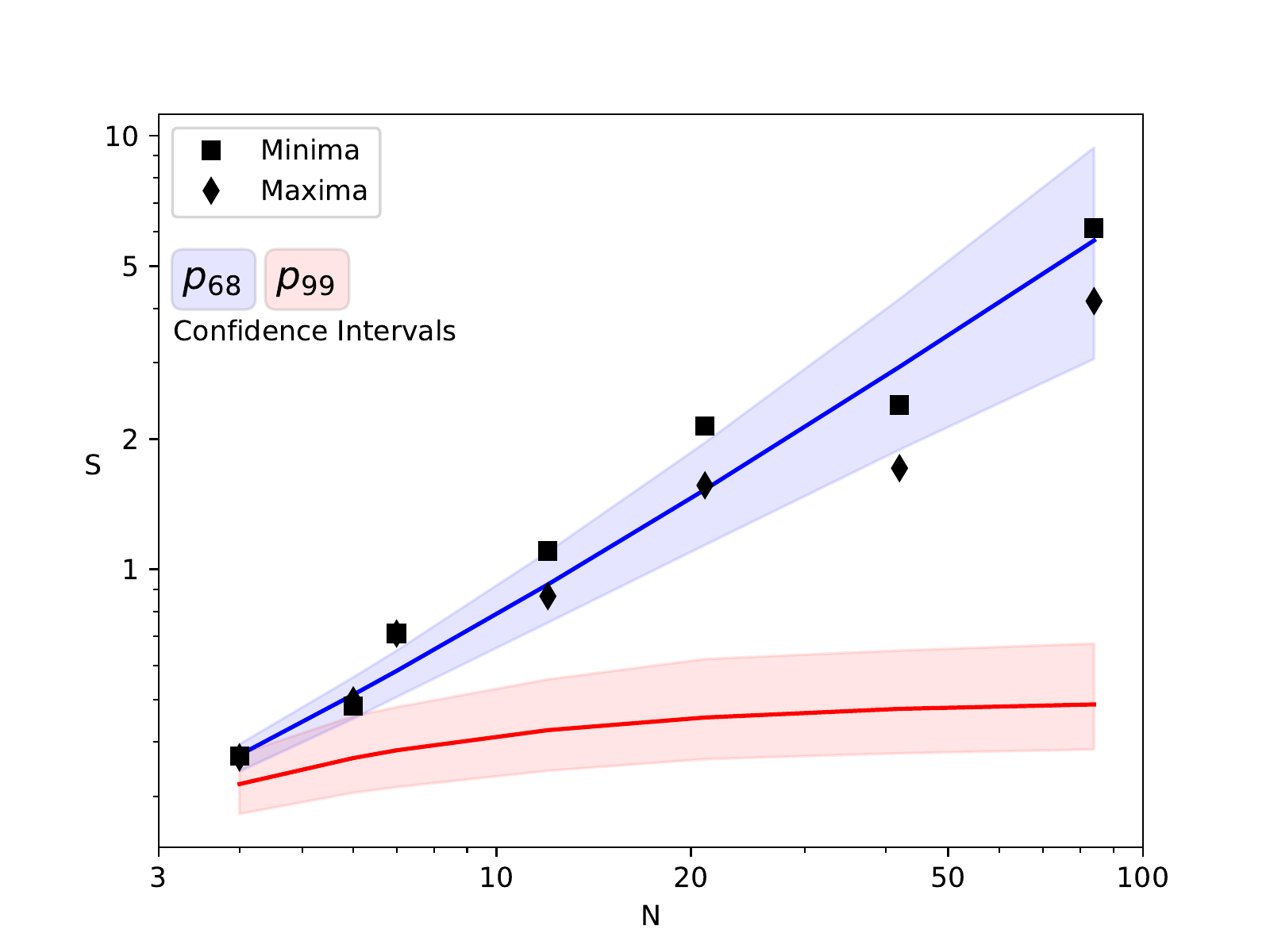}
\caption{Application of the statistical analysis to the series of 84
  cycles covering the period between the years 976 and 1888 as
  reconstructed from $^{14}$C data by \citep{Usoskin:etal:2021}. 
  {\it Upper panel:} (Corrected) Original method of \citet{Gough:1981}.
  {\it Lower panel:} Modified method \citep{Gough:1983}. The graphical
  representation of the results is the same as in the previous
  figures. The data are consistent with random walk of phase (case R)
  while, owing to the much longer data set, clock synchronization
  is rejected at levels exceeding 99\%.}
\label{fig:c14}
\end{center}
\end{figure}

The recent reconstruction of 84 cycles (between the years 976 and
1888) on the basis of the cosmogenic isotope $^{14}$C by
\citet{Usoskin:etal:2021} provides a significant extension of the database available for the analysis. Figure \ref{fig:c14} shows the results
of applying both variants of the Gough method to these data; they
confirm and significantly reinforce the results drawn from the
sunspot data: clock synchronization is now rejected at a confidence level of  at least
99\%,  while random walk remains consistent with the data.

\begin{figure}
\begin{center}
\includegraphics[width=\hsize]{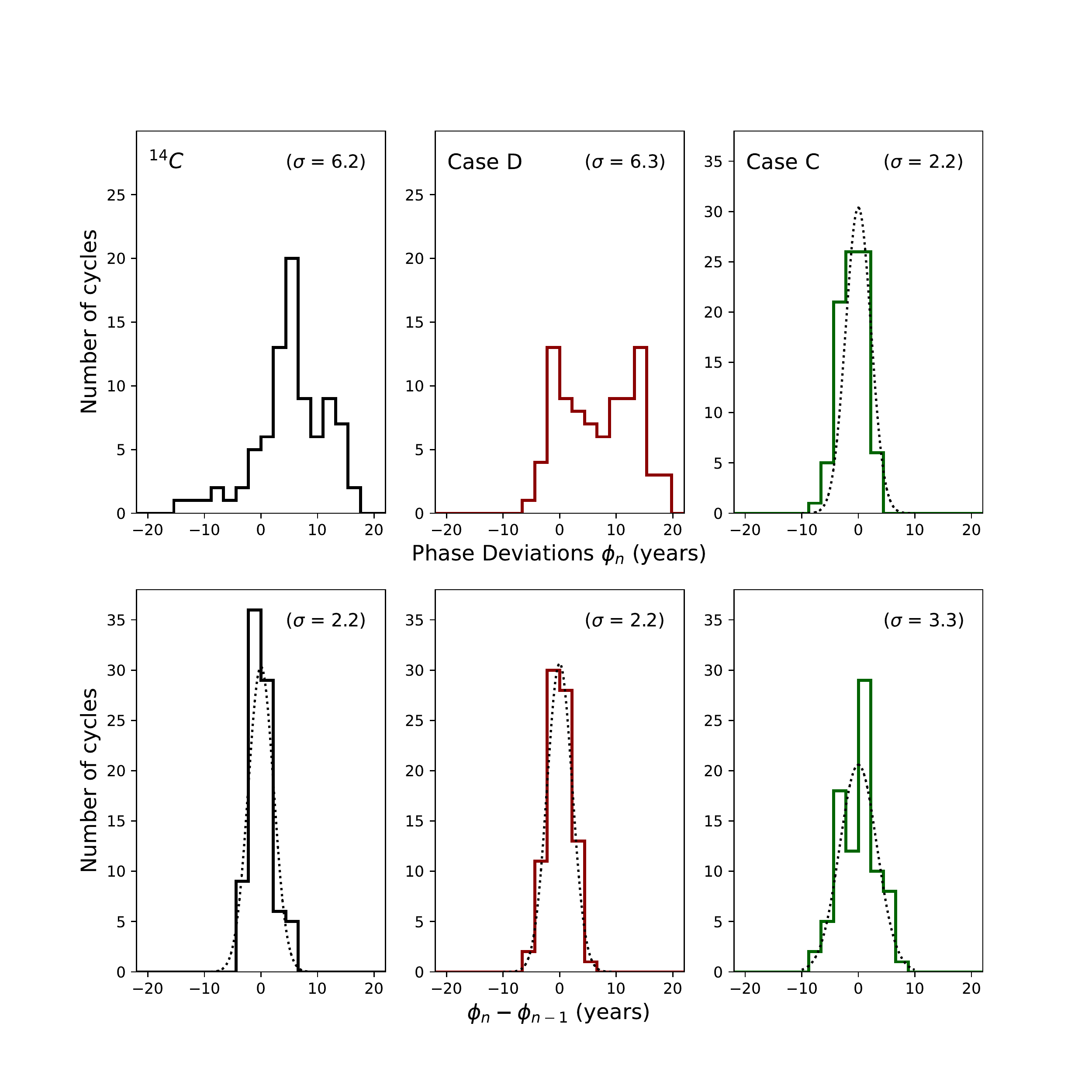}
\caption{Distributions of phase deviations (upper panels) and inferred
  phase fluctuations (differences between subsequent phase
  deviations, lower panels) for the 84 reconstructed cycles from the
  $^{14}$C data (left panels) and for simulated cycles with Gaussian
  fluctuations corresponding to case R (middle panels) and case C
  (right panels). The dotted curves indicate Gaussian fits.}
\label{fig:phase_distrib}
\end{center}
\end{figure}

The longer set of reconstructed cycles allows us to also consider the
distribution of the phase deviations, $\phi_n$, as a means to
qualitatively discriminate between the cases C and R. The upper row of
Fig.~\ref{fig:phase_distrib} shows the distribution of phase
deviations for the 84 reconstructed cycles from the $^{14}$C record
(relative to the mean period of 10.9~yr) in comparison to the
corresponding distributions from one realization of the
Monte Carlo simulations of 84 artificial cycles for case R and the same for case C. Both simulations assume a base period of 11 years with a
standard deviation of 2.2 years for the phase fluctuations.  As the
phase deviations accumulate in case R (upper middle panel), the
corresponding distribution is significantly broader ($\sigma=6.3\,$yr)
than that of case C (upper right panel, $\sigma=2.2\,$yr), the latter
reflecting the input Gaussian distribution. Likewise, the distribution
for the 84 reconstructed cycles (upper left panel, $\sigma=6.2\,$yr)
is also markedly broader than that of the simulated case C.  For case
R, we can retrieve the individual phase fluctuation, $\psi_n$, of
cycle $n$ as the difference $\phi_n - \phi_{n-1}$.  This is shown in
the lower row of Fig.~\ref{fig:phase_distrib}. For the simulated case
R, the result (based on the mean period) therefore closely reflects
the input distribution (lower middle panel).  However, if we apply the same
procedure to the simulated case C,  this necessarily leads to a
broader distribution than the input, because it results from the
difference of two Gaussian distributions, $\tau_n - \tau_{n-1}$
(lower right panel).  We can therefore use this procedure as a
qualitative method to distinguish whether a given dataset corresponds
to case R or to case C: if the distribution of $\phi_n - \phi_{n-1}$
is narrower than the original distribution, this indicates case R, while
a broader distribution favors case C.  The two panels corresponding
to the $^{14}$C reconstructions on the left side of
Fig.~\ref{fig:phase_distrib} clearly show that the reconstructed solar
cycles behave as expected for case R, that is, the distribution of the
inferred phase fluctuations for the individual cycles is significantly
narrower ($\sigma=2.2\,$yr) than that of the original phase deviations
($\sigma=6.2\,$yr).

\section{Further tests}
\label{sec_tests}

\subsection{Systematic effects}
\label{subsec_systematic}
We consider two systematic effects that can influence the epochs of
activity maxima and minima. Firstly, the ``Waldmeier effect'', which is
the correlation between the rise time from activity minimum to maximum
and the amplitude of a cycle \citep{Waldmeier:1935}, affects the times
of the maxima.  This effect enhances the variance of the period determined
from the maxima and possibly explains why the empirical $S$ values
corresponding to the set of cycle maxima tend to be somewhat lower
than those for the minima (see
Figs.~\ref{fig:sunspots_24}--\ref{fig:c14}).  Following
\citet{Dicke:1978}, \citet{Gough:1981, Gough:1983} corrected the
sunspot data for the Waldmeier effect and found that the results were
not significantly affected.

\begin{figure}
\begin{center}
\includegraphics[width=\hsize]{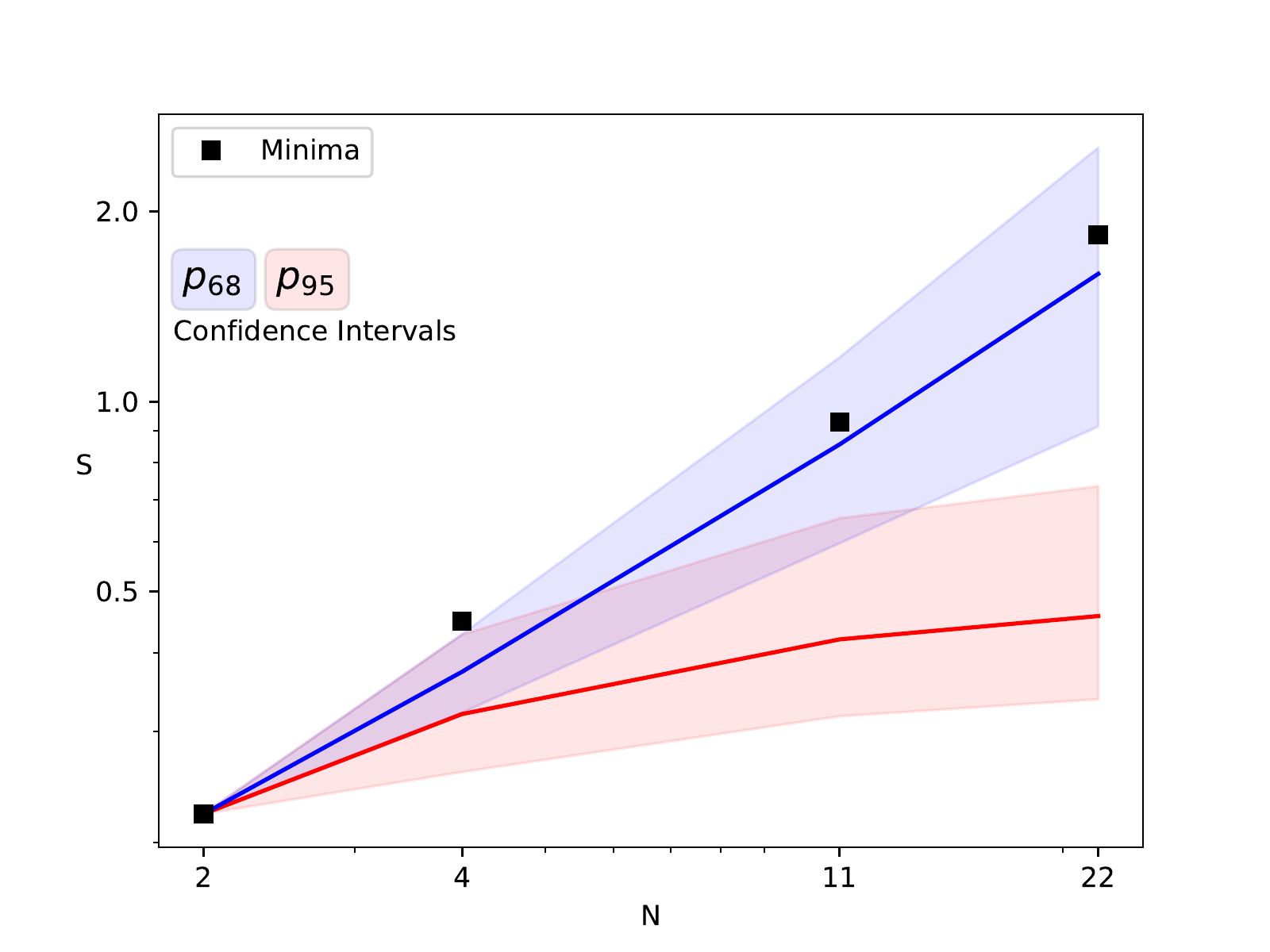}
\caption{Application of the modified method of \citet{Gough:1983} to
  the unbiased starting dates of solar cycles 1 to 23 determined by
  \citet{Hathaway:etal:1994} and \citet{Hathaway:2011}.  }
\label{fig:hathaway}
\end{center}
\end{figure}

Secondly, activity cycles overlap, that is, flux emergence of a new cycle
already starts at mid solar latitudes while the previous cycle
is still present at low latitudes.  In combination with the Waldmeier
effect, this leads to an amplitude-dependent shift of the minimum
epochs \citep{Cameron:Schuessler:2007}. Incidentally, this shift also
explains the empirical statistical relationship between cycle length
and amplitude. \citet{Hathaway:etal:1994} and \citet{Hathaway:2011}
used curve fitting to an empirical standard cycle shape to obtain starting dates of sunspot cycles 1 to 23 (1755 until 1996) that are
not affected by cycle overlap. The result of applying the modified
Gough test (Sect.~\ref{subsec_Gough_1983}) to these data is given in
Fig.~\ref{fig:hathaway}. Comparison with the square symbols
presented in Figs.~\ref{fig:sunspots_24}b and \ref{fig:sunspots_28}b
shows that the shift of the minimum epochs due to cycle overlap and the
Waldmeier effect does not significantly affect the statistical results
of the Gough test.

\subsection{Dependence on the length of the data set}
\label{subsec_FAP}
The length of the data set, that is, the number of observed cycles, is
crucial for the capability of a statistical test to distinguish
between case C and case R. Our results show that the set of 84 cycles
reconstructed from the $^{14}$C record permits a much more stringent
answer than the 28 cycles provided by the sunspot record. In order to
systematically assess the reliability of the original and modified
Gough methods, we used Monte Carlo simulations to determine the
false-alarm probability (FAP) with respect to the 99\% confidence
interval as a function of the total number of cycles, $M$.  For each
given $M$, we computed $10^6$ realizations of each of the models, that is,
case C (Eq.~\ref{eq:period_clock}) and case R
(Eq.~\ref{eq:period_dynamo}).  For each realization, we calculated the
value of our statistic, $S$, for $N=M$, which provides the largest
difference between $S_{\mathrm R}$ and $S_{\mathrm C}$, and is therefore
the most sensitive. We then determined the FAPs, that is, the percentage of
the realizations of case C with $S$ values within the 99\% confidence
interval for case R and, conversely, the percentage of the
realizations of case R with $S$ values within the 99\% confidence
interval for case C.

\begin{figure}
\begin{center}
\includegraphics[width=\hsize]{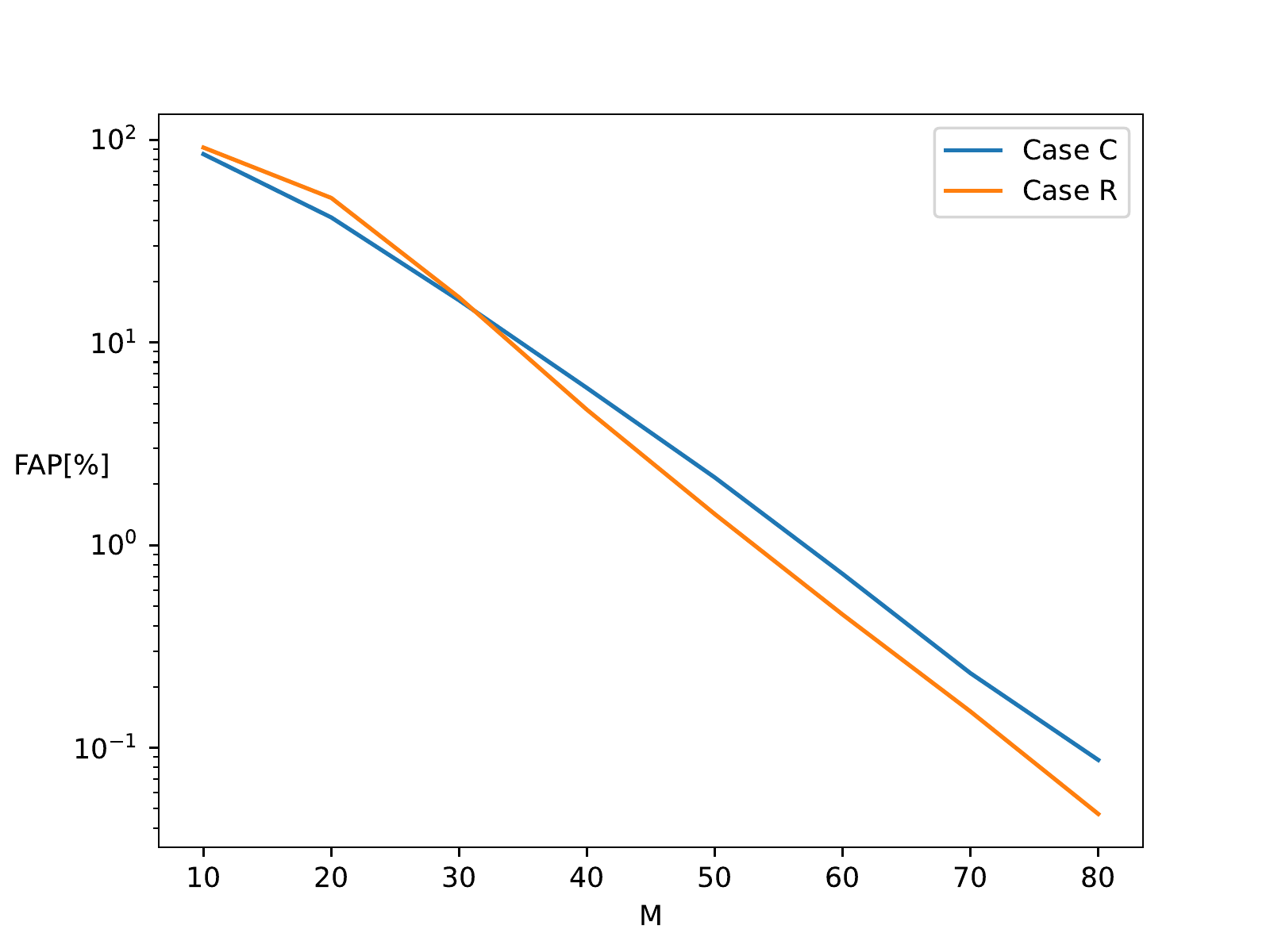}
\includegraphics[width=\hsize]{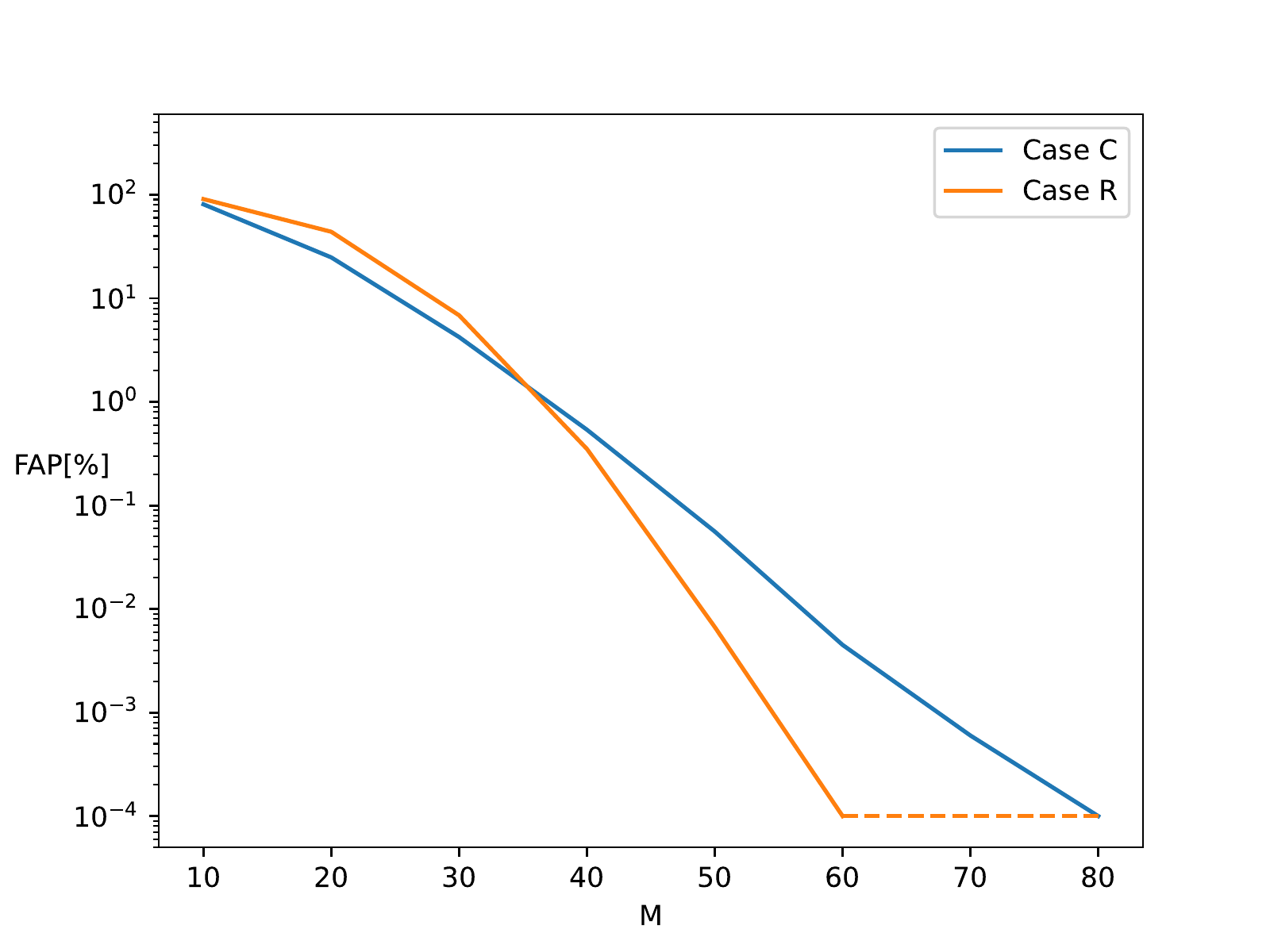}
\caption{False-alarm probabilities as a function of the total number
  of cycles in the data set, based on Monte Carlo simulations.  {\it
    Upper panel:} (Corrected) Original method of \citet{Gough:1981}.
  {\it Lower panel:} Modified method \citep{Gough:1983}.  The red
  lines give the FAP that a realization of case C falls within the
  99\% confidence interval for case R, and vice versa for the blue
  lines. The dashed line segment indicates an upper limit.}
\label{fig:FAP}
\end{center}
\end{figure}

The results of this procedure for both the original and the modified
Gough test are shown in Fig.~\ref{fig:FAP}. The red curves correspond
to the FAP for case R, meaning the probability that a realization of
case R would fall within the 99\% confidence interval of case C, or in other words
the probability of misclassification as case C at that
level. Likewise, the blue lines give the probability for misclassfying
a realization of case C as case R. The figures clearly show that the
modified method provides significantly lower values of the FAP, particularly
for larger $M$.  Reasonably low values of less than 1\% FAP are
reached for $M>50$ for the original Gough test and $M>30$ for the
modified test. This corresponds to the results shown in
Sect.~\ref{sec_results}: while the clock hypothesis can be safely
rejected at the 99\% significance level for $M=84$ using the
$^{14}$C-based data, this can only be done at 95\% significance level for
$M=24(28)$ on the basis of the sunspot record.

\subsection{The method of Eddington and Plakidis}
\label{subsec_Eddington}

A problem very similar to the question of phase stability of the solar
cycle was considered by \citet{Eddington:Plakidis:1929}. These authors considered the question of whether the brightness changes of
long-period variable stars have an underlying fixed period with
superposed observational errors (analogous to the clock
case in the context of the present study) or if the individual periods vary randomly without memory (i.e., the random walk case). The authors considered
the mean value, $\bar{u}_x$, of the accumulated phase error after $x$
epochs of maximum brightness with respect to the mean period over the
whole length of the data set. Assuming that the period variations
consist of a combination of accumulating (random walk) and
nonaccumulating (clock-type) fluctuations, the expectation value is
given by
\begin{equation}
\bar{u}_x = \sqrt{2\alpha^2 + \epsilon^2 x \left( 1-{x\over M} \right)}
\label{eq:eddington}
,\end{equation}
where $\alpha$ is the amplitude of the synchronized fluctuations (e.g.,
measurement errors) and $\epsilon$ the amplitude of the cumulating
fluctuations (leading to random walk of phase).  A correction factor
(in parentheses) is applied to account for the cases where $x$ becomes a
considerable fraction of the total number of periods, $M$, in the data
set. $\bar{u}_x$ can be calculated from the data and a
least-squares fit to Eq.~\ref{eq:eddington} provides empirical values
of the amplitudes $\alpha$ and $\epsilon$, from which the relative
importance of clock and random walk can be assessed.

\begin{figure}
\begin{center}
\includegraphics[width=\hsize]{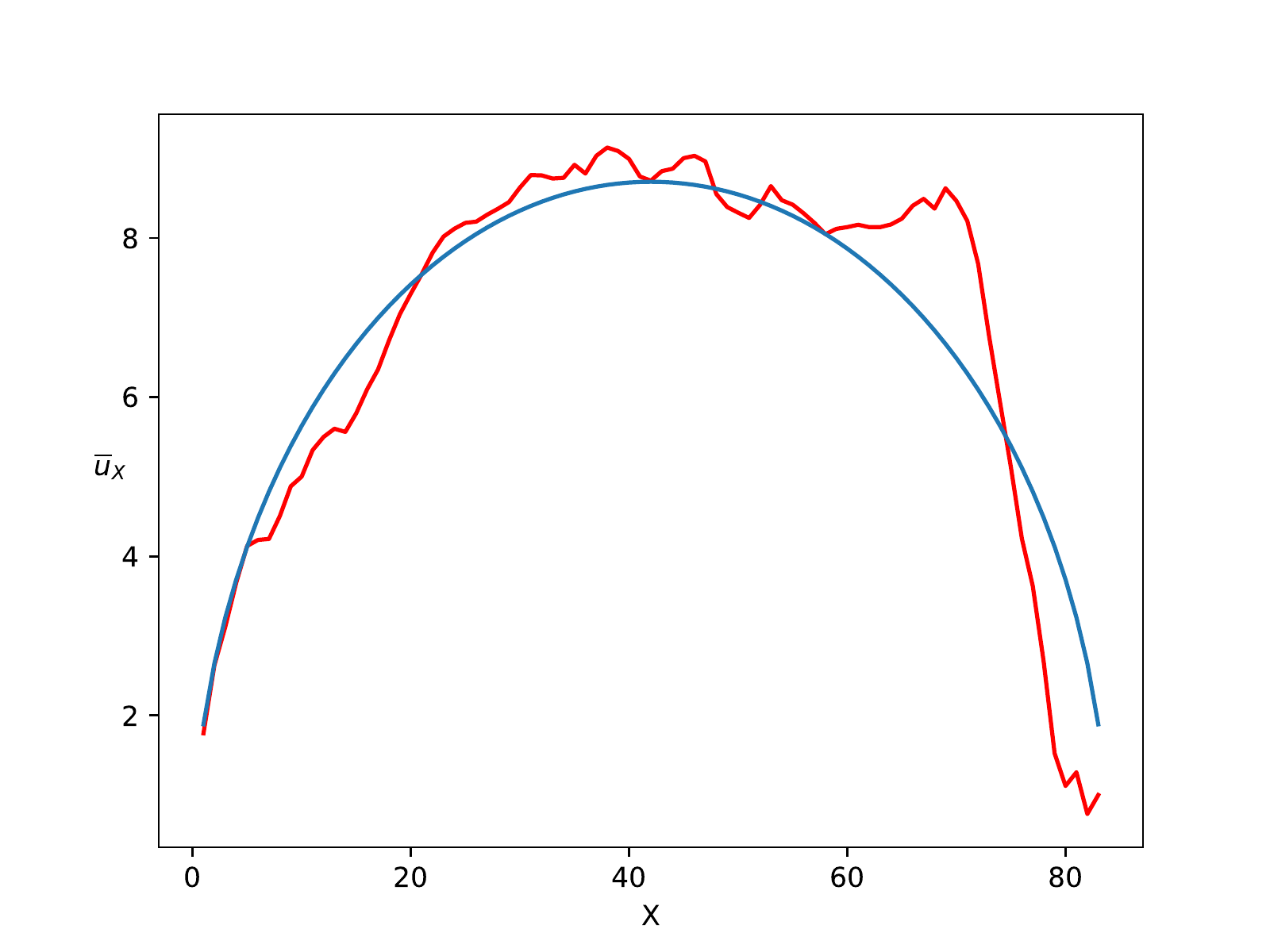}
\caption{Application of the test of \citet{Eddington:Plakidis:1929} to
  the 84 reconstructed cycles on the basis of the $^{14}$C data. Shown
  is the mean accumulated phase error after $x$ cycles as a function
  of $x$. The red curve results from the activity minima of the
  reconstructed cycles, the blue curve is the corresponding fit to the
  theoretical relation given by Eq.~\ref{eq:eddington} for $x \leq
  50$. The resulting fit parameters clearly indicate that the system
  shows random walk of phase.}
\label{fig:eddington}
\end{center}
\end{figure}

We calculated $\bar{u}_x$ on the basis of the minimum epochs of
the 84 activity cycles determined from the $^{14}$C data. The results
and the fit to Eq.~\ref{eq:eddington} are shown in
Fig.~\ref{fig:eddington}. As the individual values of the
accumulated phase, $u_x$, are significantly affected by overlap for
large values of $x$, we restricted the fit to $x\leq 50$. The fit
results in $\alpha<10^{-4}\,$yr for the amplitude of the
nonaccumulating fluctuations and $\epsilon=1.9\,$yr for the amplitude
of the cumulating fluctuations. This clearly demonstrates that the
system is nearly exclusively governed by random walk of the cycle phase, thus
corroborating the results of the Gough test.

\section{Conclusions}
\label{sec_concl}
Using rigorous statistical tests, we investigated whether or not the
phase of the solar cycle is stable, that is, whether it is
synchronized by some kind of clock or follows a random walk. Our
analysis is based on 28 activity cycles from the historical sunspot
record and 84 cycles reconstructed from $^{14}$C data in tree rings.
The tests reveal that these data unequivocally favor random walk of the
cycle phase and exclude clock synchronization, with high significance
levels (95\% for sunspot data and greatly exceeding 99\% for the
$^{14}$C data). Consequently, data that unambiguously correspond to
solar activity cycles provide no basis for the ``planetary
hypothesis'' or for hypothetical high-Q oscillations in the radiative
interior as drivers or ``synchronizers'' of the solar cycle.  Likewise,
data for an exoplanetary system also provide no evidence for planetary
synchronization of stellar cycles \citep{Obridko:etal:2022}. The
consistency of the phase evolution with random walk suggests that a randomly
perturbed, memoryless dynamo action is the cause of the solar activity
cycle \citep[e.g.,][]{Charbonneau:Dikpati:2000,
  Cameron:Schuessler:2017, Karak:Miesch:2018}.

\begin{acknowledgements}
Ilya Usoskin kindly provided the reconstructed sunspot numbers
obtained by \citet{Usoskin:etal:2021} in electronic form. We are
grateful to the referee for making us aware of the paper by
\citet{Eddington:Plakidis:1929}.
\end{acknowledgements}

\bibliographystyle{aa}
\bibliography{44997.bbl}

\begin{thebibliography}{35}
\expandafter\ifx\csname natexlab\endcsname\relax\def\natexlab#1{#1}\fi

\bibitem[{{Brehm} {et~al.}(2021){Brehm}, {Bayliss}, {Christl}, {Synal},
  {Adolphi}, \& {Etal}}]{Brehm:etal:2021}
{Brehm}, N., {Bayliss}, A., {Christl}, M., {et~al.} 2021, Nature Geoscience,
  14, 10

\bibitem[{{Cameron} \& {Sch{\"u}ssler}(2007)}]{Cameron:Schuessler:2007}
{Cameron}, R.~H. \& {Sch{\"u}ssler}, M. 2007, \apj, 659, 801

\bibitem[{{Cameron} \& {Sch{\"u}ssler}(2017)}]{Cameron:Schuessler:2017}
{Cameron}, R.~H. \& {Sch{\"u}ssler}, M. 2017, \apj, 843, 111

\bibitem[{{Carrasco} {et~al.}(2020){Carrasco}, {Gallego}, {Arlt}, \&
  {Vaquero}}]{Carrasco:etal:2020}
{Carrasco}, V.~M.~S., {Gallego}, M.~C., {Arlt}, R., \& {Vaquero}, J.~M. 2020,
  \apj, 904, 60

\bibitem[{{Charbonneau} \& {Dikpati}(2000)}]{Charbonneau:Dikpati:2000}
{Charbonneau}, P. \& {Dikpati}, M. 2000, \apj, 543, 1027

\bibitem[{{Dicke}(1970)}]{Dicke:1970}
{Dicke}, R.~H. 1970, in IAU Colloq. 4: Stellar Rotation, ed. A.~{Slettebak}
  (Reidel, Dordrecht), 289

\bibitem[{{Dicke}(1978)}]{Dicke:1978}
{Dicke}, R.~H. 1978, \nat, 276, 676

\bibitem[{{Eddington} \& {Plakidis}(1929)}]{Eddington:Plakidis:1929}
{Eddington}, A.~S. \& {Plakidis}, S. 1929, \mnras, 90, 65

\bibitem[{{Gough}(1978)}]{Gough:1978}
{Gough}, D. 1978, in Pleins Feux sur la Physique Solaire, ed. S.~{Dumont} \&
  J.~{Roesch}, 81

\bibitem[{{Gough}(1981)}]{Gough:1981}
{Gough}, D. 1981, in NASA Conference Publication, Vol. 2191, 185--206

\bibitem[{{Gough}(1983)}]{Gough:1983}
{Gough}, D.~O. 1983, ESA Journal, 7, 325

\bibitem[{{Gough}(1988)}]{Gough:1988}
{Gough}, D.~O. 1988, in Solar-Terrestrial Relationships and the Earth
  Environment in the last Millennia, ed. G.~Cini~Castagnoli (North-Holland,
  Amsterdam etc.), 90

\bibitem[{{Hathaway}(2011)}]{Hathaway:2011}
{Hathaway}, D.~H. 2011, \solphys, 273, 221

\bibitem[{{Hathaway} {et~al.}(1994){Hathaway}, {Wilson}, \&
  {Reichmann}}]{Hathaway:etal:1994}
{Hathaway}, D.~H., {Wilson}, R.~M., \& {Reichmann}, E.~J. 1994, \solphys, 151,
  177

\bibitem[{{Hoyng}(1996)}]{Hoyng:1996}
{Hoyng}, P. 1996, \solphys, 169, 253

\bibitem[{{Karak} \& {Miesch}(2018)}]{Karak:Miesch:2018}
{Karak}, B.~B. \& {Miesch}, M. 2018, \apjl, 860, L26

\bibitem[{{Lomb}(2013)}]{Lomb:2013}
{Lomb}, N. 2013, in Journal of Physics Conference Series, Vol. 440, 012042

\bibitem[{{Nataf}(2022)}]{Nataf:2022}
{Nataf}, H.-C. 2022, \solphys, 297, 107

\bibitem[{{Obridko} {et~al.}(2022){Obridko}, {Katsova}, \&
  {Sokoloff}}]{Obridko:etal:2022}
{Obridko}, V.~N., {Katsova}, M.~M., \& {Sokoloff}, D.~D. 2022, arXiv e-prints,
  arXiv:2208.06190

\bibitem[{{Russell} {et~al.}(2019){Russell}, {Jian}, \&
  {Luhmann}}]{Russell:etal:2019}
{Russell}, C.~T., {Jian}, L.~K., \& {Luhmann}, J.~G. 2019, Rev. of Geophys.,
  57, 1129

\bibitem[{{Schove}(1955)}]{Schove:1955}
{Schove}, D.~J. 1955, \jgr, 60, 127

\bibitem[{{Schove}(1983)}]{Schove:1983}
{Schove}, D.~J. 1983, {Sunspot cycles} (Hutchinson Ross Publ. Comp., Benchmark
  Papers in Geology, Vol.68))

\bibitem[{{Stefani} {et~al.}(2020){Stefani}, {Beer}, {Giesecke}, {Gloaguen},
  {Seilmayer}, {Stepanov}, \& {Weier}}]{Stefani:etal:2020}
{Stefani}, F., {Beer}, J., {Giesecke}, A., {et~al.} 2020, Astronomische
  Nachrichten, 341, 600

\bibitem[{{Stefani} {et~al.}(2021){Stefani}, {Stepanov}, \&
  {Weier}}]{Stefani:etal:2021}
{Stefani}, F., {Stepanov}, R., \& {Weier}, T. 2021, \solphys, 296, 88

\bibitem[{{Stephenson}(1988)}]{Stephenson:1988}
{Stephenson}, F.~R. 1988, in Solar-Terrestrial Relationships and the Earth
  Environment in the last Millennia, ed. G.~Cini~Castagnoli (North-Holland,
  Amsterdam etc.), 133

\bibitem[{{Stephenson}(1990)}]{Stephenson:1990}
{Stephenson}, F.~R. 1990, Phil. Trans. Roy. Soc. London, Series A, 330, 499

\bibitem[{{Stix}(1983)}]{Stix:1983}
{Stix}, M. 1983, Mitteilungen der Astronomischen Gesellschaft Hamburg, 60, 95,
  https://ui.adsabs.harvard.edu/abs/1983MitAG..60...95S

\bibitem[{{Stix}(1984)}]{Stix:1984}
{Stix}, M. 1984, Astronomische Nachrichten, 305, 215

\bibitem[{{Usoskin} {et~al.}(2015){Usoskin}, {Arlt}, {Asvestari}, {Hawkins},
  {K{\"a}pyl{\"a}}, {Kovaltsov}, {Krivova}, {Lockwood}, {Mursula}, {O'Reilly},
  {Owens}, {Scott}, {Sokoloff}, {Solanki}, {Soon}, \&
  {Vaquero}}]{Usoskin:etal:2015}
{Usoskin}, I.~G., {Arlt}, R., {Asvestari}, E., {et~al.} 2015, \aap, 581, A95

\bibitem[{{Usoskin} {et~al.}(2021){Usoskin}, {Solanki}, {Krivova}, {Hofer},
  {Kovaltsov}, {Wacker}, {Brehm}, \& {Kromer}}]{Usoskin:etal:2021}
{Usoskin}, I.~G., {Solanki}, S.~K., {Krivova}, N.~A., {et~al.} 2021, \aap, 649,
  A141

\bibitem[{{Vaquero} \& {V{\'a}zquez}(2009)}]{Vaquero:Vazquez:2009}
{Vaquero}, J.~M. \& {V{\'a}zquez}, M. 2009, Astrophysics and Space Science
  Library, Vol. 361, {The Sun Recorded Through History} (Springer)

\bibitem[{{Waldmeier}(1935)}]{Waldmeier:1935}
{Waldmeier}, M. 1935, Astronomische Mitteilungen der Eidgen{\"o}ssischen
  Sternwarte Z{\"u}rich, 14, 105

\bibitem[{{Wittmann}(1978)}]{Wittmann:1978}
{Wittmann}, A. 1978, \aap, 66, 93

\bibitem[{{Wittmann} \& {Xu}(1987)}]{Wittmann:Xu:1987}
{Wittmann}, A.~D. \& {Xu}, Z.~T. 1987, \aaps, 70, 83

\bibitem[{{Yule}(1927)}]{Yule:1927}
{Yule}, G.~U. 1927, Phil. Trans. Roy. Soc. London A, 226, 267

\end{thebibliography}


\begin{appendix}
\onecolumn

\section{Calculation of the statistic $S_R$}
\setcounter{equation}{0}
\renewcommand{\theequation}{A.\arabic{equation}}
In what follows, we describe the calculation of the statistic 
$S = E\{\sigma_\Phi^2\}/E\{\sigma_{\mathrm P}^2\}$ for case R (random walk
of phase) as would be expected for a randomly perturbed dynamo.
We follow \citet{Gough:1981} and write, for the $n^{\rm th}$ observed period, 
\begin{equation}
  P_{n} = P_{\mathrm R} + \psi_n \,, 
\label{eq:app_period_dynamo}
\end{equation}
where $P_{\mathrm R}$ is the ``true'' period (in the absence of fluctuations)
and $\psi_n$ denotes random, uncorrelated perturbations with zero mean
and amplitude $\Psi$, namely
\begin{equation}
  E\{\psi_n\} = 0 \;\;\; \mathrm{and} \;\;\; E\{\psi_n\psi_{n'}\} = \Psi^2\delta_{nn'} 
 \,.
\label{eq:app_psi_expect}
\end{equation}
The unknown period $P_{\mathrm R}$ is estimated via the arithmetic mean of
the periods given by
\begin{equation}
  \langle P \rangle = {1\over N} \sum_{i=1}^N P_{i} = P_{\mathrm R} + {1\over N} \sum_{i=1}^N \psi_i \,. 
\label{eq:app_meanp}
\end{equation}
For the expectation value of the variance of period, we obtain
\begin{eqnarray}
  E\left\{\sigma{_{\mathrm P}}^2\right\} &=& E\left\{ {1\over N}  \sum_{i=1}^N \left( P_i - \langle P \rangle\right)^2 \right\} 
                    = E\left\{ {1\over N}  \sum_{i=1}^N \left(\psi_i - {1\over N} \sum_{j=1}^N \psi_j\right)^2\right\} =  \nonumber \\
                    &=& E\left\{ {1\over N}  \sum_{i=1}^N \left[ \psi_i^2 - {2\psi_i\over N}\sum_{j=1}^N \psi_j
                                       + {1\over N^2}\sum_{k=1}^N\sum_{\ell=1}^N \psi_k\psi_\ell \right]\right\} 
                    = \Psi^2 - {2\Psi^2\over N} + {\Psi^2\over N} = {N-1\over N}\Psi^2 \;.   
\label{eq:app_sigp}
\end{eqnarray}
The phase deviations are defined by $\phi_n = t_n - t_0 - n\cdot\langle P \rangle$, where
\begin{equation}
  t_n = n\cdot P_{\mathrm R} +  \sum_{i=0}^n\psi_i    
\label{eq:app_tn}
,\end{equation}
and so we obtain
\begin{equation}
  \phi_{n} = t_n - T_{n} =  \sum_{i=1}^n \psi_i -  {n\over N} \sum_{i=1}^N \psi_i \,. 
\label{eq:app_phases}
\end{equation}
We now consider the expectation value of the variance of the phase deviation,
\begin{equation}
  E\left\{ \sigma_\phi^2\right\} = E\left\{ {1\over N+1} \sum_{i=0}^N\phi_i^2 
           - \left({1\over N+1}  \sum_{i=0}^N\phi_i \right)^2 \right\} \equiv {\rm I} - {\rm II}  \,,  
\label{eq:app_variance_phase}
\end{equation}
which consists of two terms (denoted by I and II). For calculating term I, we first consider
\begin{eqnarray}
  E\left\{\phi_n^2\right\} &=& E\left\{ \sum_{i=1}^n \sum_{j=1}^n \psi_i\psi_j - {2n\over N} \sum_{i=1}^n\psi_i \sum_{j=1}^N\psi_j
                                 + {n^2\over N^2} \sum_{i=1}^N \sum_{j=1}^N \psi_i\psi_j    \right\} =  \nonumber \\
                    &=& n\Psi^2 - {2n^2\over N}\Psi^2 + {n^2\over N}\Psi^2 = n\left(1 - {n\over N}\right) \Psi^2 \;,   
\label{eq:app_exp_phi2}
\end{eqnarray}
which leads to
\begin{equation}
  {\rm I}= {1\over N+1} \sum_{i=0}^N E\left\{ \phi_i^2\right\} = {\Psi^2\over N+1} \sum_{i=0}^N\left(i-{i^2\over N}\right)\,.  
\label{eq:app_term_Ia}
\end{equation}
With 
\begin{equation}
  \sum_{i=0}^N i = {N(N+1)\over 2} \;\; {\rm and} \;\;   \sum_{i=0}^N i^2 = {N(N+1)(2N+1)\over 6} 
\label{eq:app_sums}
,\end{equation}
we obtain
\begin{equation}
   {\rm I} = \left( {N\over 2} - {2N+1\over 6} \right) \Psi^2 = \left( {N-1\over 6} \right) \Psi^2 \,. 
\label{eq:app_term_I}
\end{equation}
This agrees with the result given by \citet{Gough:1981} for the
complete variance of the phase deviation.  However, for the correct
result, we must also consider the second term, II, in
Eq.~\ref{eq:app_variance_phase}, namely
\begin{eqnarray}
  {\rm II} &=& E\left\{ \left({1\over N+1}  \sum_{i=0}^N\phi_i \right)^2 \right\} 
           = {1\over (N+1)^2} E\left\{ \sum_{i,j=0}^N \phi_i\phi_j \right\} = \nonumber \\
           &=& {1\over (N+1)^2} E\left\{ \sum_{i,j=0}^N \left(\sum_{k=1}^i\psi_k-{i\over N}\sum_{\ell=1}^N\psi_\ell\right)
               \left( \sum_{m=1}^j\psi_m-{j\over N}\sum_{n=1}^N\psi_n\right)\right\} = \nonumber \\
           &=& {1\over (N+1)^2} E\left\{ \sum_{i,j=0}^N \left[ \sum_{k=1}^i\sum_{m=1}^j\psi_k\psi_m
               + {ij\over N^2}\sum_{\ell=1}^N\sum_{n=1}^N\psi_\ell\psi_n - {2i\over N}\sum_{\ell=1}^N\sum_{m=1}^j\psi_\ell\psi_m \right]\right\} \,.
\label{eq:app_term_IIa}
\end{eqnarray}
We have
\begin{equation}
    E\left\{ \sum_{k=1}^i\sum_{m=1}^j\psi_k\psi_m \right\} = \Psi^2  \sum_{k=1}^i\sum_{m=1}^j \delta_{km} = \Psi^2  \min (i,j) \,, 
\label{eq:app_term_IIb}
\end{equation}
which means that for the first term in curly brackets in the bottom row of Eq.~(\ref{eq:app_term_IIa}) we obtain
\begin{eqnarray}
   E\left\{ \sum_{i,j=0}^N \sum_{k=1}^i\sum_{m=1}^j\psi_k\psi_m \right\} 
         &=& \Psi^2  \sum_{i,j=0}^N \min (i,j)  = \Psi^2 \left( 2\sum_{j=1}^N\sum_{i=0}^{j-1}i + \sum_{i=0}^N i \right) = \nonumber \\
         &=& \Psi^2 \left( 2\sum_{j=1}^N {(j-1)j\over 2} + \sum_{i=0}^N i \right) = \Psi^2 \sum_{j=1}^N j^2 
         = {N(N+1)(2N+1)\over 6} \Psi^2  
\label{eq:app_term_IIc}
.\end{eqnarray}
For the second term in curly brackets in the bottom row of Eq.~(\ref{eq:app_term_IIa}), we find
\begin{equation}
    E\left\{ \sum_{i,j=0}^N {ij\over N^2}\sum_{\ell=1}^N\sum_{n=1}^N\psi_\ell\psi_n  \right\} = \Psi^2 \sum_{i,j=0}^N  {ij\over N} \,, 
\label{eq:app_term_IId}
\end{equation}
while the third term yields
\begin{equation}
   E\left\{ \sum_{i,j=0}^N \left( -{2i\over N} \right) \sum_{\ell=1}^N\sum_{m=1}^j\psi_\ell\psi_m \right\} 
          = \Psi^2 \sum_{i,j=0}^N \left( -{2i\over N} \right) \sum_{\ell=1}^N\sum_{m=1}^j \delta_{\ell m} 
          = \Psi^2  \sum_{i,j=0}^N \left( -{2ij\over N} \right) \,.  
\label{eq:app_term_IIe}
\end{equation}
This means that the sum of the second and the third term becomes
\begin{equation}
   - {\Psi^2\over N} \sum_{i,j=0}^N ij = - {\Psi^2\over N} \sum_{i=0}^N i \sum_{j=0}^N j = - {N(N+1)^2\over 4}\Psi^2    \,.  
\label{eq:app_term_IIf}
\end{equation}
Adding the terms in Eqs.~\ref{eq:app_term_IIc} and \ref{eq:app_term_IIf}, we obtain
\begin{eqnarray}
   {\rm {II}} = E\left\{ \left({1\over N+1}  \sum_{i=0}^N\phi_i \right)^2 \right\} 
              &=& {1\over (N+1)^2} \left[ {N(N+1)(2N+1)\over 6} - {N(N+1)^2\over 4}\right] \Psi^2 = 
               {N(N-1)\over 12(N+1)} \Psi^2 \,.
\label{eq:app_term_II_res}
\end{eqnarray}
Using Eqs.~(\ref{eq:app_term_I}) and (\ref{eq:app_term_II_res}), the result for the variance of the phase deviation becomes
\begin{equation}
  E\left\{ \sigma{_\phi}^2\right\} = {\rm I} - {\rm II} = {(N+2)(N-1)\over 12(N+1)} \Psi^2  \,.  
\label{eq:app_variance_phase_res}
\end{equation}
With Eq.~\ref{eq:app_sigp}, we finally determine the statistic $S_{\mathrm R}$ for the case of random phase walk, that is,
\begin{equation}
  S_{\mathrm R} = {E\left\{ \sigma_\phi^2\right\}\over E\left\{ \sigma_{\mathrm P}^2\right\}}  =  {N(N+2)\over 12(N+1)} \,.  
\label{eq:app_SD}
\end{equation}
For $N\to\infty$, $S_{\mathrm R}$ behaves as $N/12$, in contrast to $N/6$ as given by \citet{Gough:1981}. 

\end{appendix}
\end{document}